\documentclass{wqcd03}                 

\usepackage{txfonts}                   
\confname{QCD@Work 2003 - International Workshop on QCD, Conversano, Italy, 
14--18 June 2003}

\title{Scalar mesons and $\delta_{0}-\delta_{2}$ at KLOE}

\author{The KLOE collaboration~\thanks{The KLOE collaboration: 
  A.~Aloisio, F.~Ambrosino, A.~Antonelli, 
  M.~Antonelli, C.~Bacci, G.~Bencivenni, S.~Bertolucci, C.~Bini, 
  C.~Bloise, V.~Bocci
% ...
}   \\ 
  Presented by Claudio Gatti}
\address{Dipartimento di Fisica, Universit\`a di Roma ``La Sapienza''
and INFN Sezione di Roma, Italy}

\begin{document}
  
  \begin{abstract} The KLOE measurements of $\phi$ radiative decays and of 
    the ratio 
    $\Gamma(K_{S}\to \pi^{+}\pi^{-}(\gamma))/\Gamma(K_{S}\to \pi^{0}\pi^{0})$
    are discussed. The first measurement aims at understanding the real nature of
    the scalar mesons $a_{0}$ and $f_{0}$, whose possible compositions are 
    $q\bar{q}$, $qq\bar{q}\bar{q}$, or $K\bar{K}$ molecule. The second measurement
    is related to the estimate of the phase shift difference
    $\delta_{0}-\delta_{2}$. Previous measurements were affected
    by a large uncertainty due to the presence of radiative effects which are 
    correctly included in our analysis.
  \end{abstract}

\maketitle


\section{The KLOE experiment}
\begin{figure}[h]
\hbox to\hsize{\hss
\includegraphics[width=\hsize]{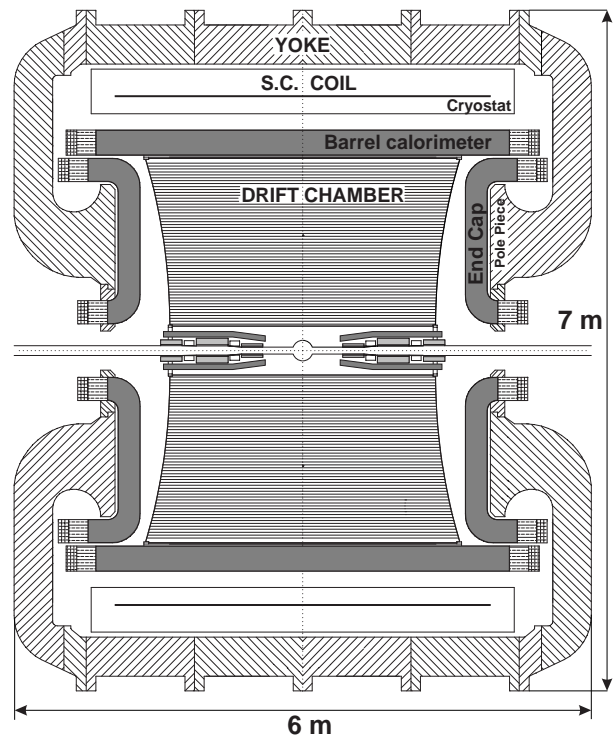}
\hss}
\caption{Section of the KLOE detector.}
\label{kloesec}
\end{figure}

The KLOE experiment collects data at DA$\Phi$NE, an 
$e^{+} e^{-}\mbox{-collider}$ with center of mass energy 
$\sim 1020~\mbox{MeV}$ corresponding to the $\phi$ meson mass. 
The detector consists of a large drift chamber surrounded by an 
electromagnetic calorimeter, both embedded in
a magnetic field of $\sim 0.5~\mbox{Tesla}$ (Fig.~\ref{kloesec}).

The drift chamber~\cite{dckloe} is a cylinder 4 m in diameter 
and 3.3 m in length, with a stereo write-geometry for the reconstruction of the
longitudinal coordinate. The momentum resolution for particles with 
$p \geq 100~\mbox{MeV}$ is $\sigma_{p}/p\leq 0.5\%$.
The calorimeter~\cite{emckloe} is a lead/scintillating-fiber 
sampling calorimeter, 
composed of a barrel and two end-caps. The energy resolution is 
$\sigma_{E}/E = 5.7\%/\sqrt{E(\mbox{GeV})}$. The intrinsic time resolution is
$\sigma_{t} = 54~\mbox{ps}/\sqrt{E(\mbox{GeV})} \oplus 50~\mbox{ps}$.

KLOE collected $\sim 500~\mbox{pb}^{-1}$ in three years of data taking.
During the year 2002 DA$\Phi$NE reached a maximum luminosity of 
$\sim 8 \times 10^{31} \mbox{cm}^{-2}\mbox{s}^{-1}$. 

\section{Scalar mesons}

SU(3) flavor symmetry has been very useful in classifying mesons and
baryons. Both the pseudoscalar ($J^{PC}=0^{-+}$) and vector 
mesons ($J^{PC}=1^{--}$) clearly show the nonet structure
expected for $q\bar{q}$ states, with masses following expectation.
The same is not true for scalar mesons ($J^{PC}=0^{++}$). 
In fact, instead of the expected
9 particles, 15 scalar mesons are observed below 2 GeV as listed
by the Particle Data Group~\cite{PDG}, with masses not in accord
with the $q\bar{q}$ model.
The excess of particles can
be explained by introducing non-conventional mesons such as
glue balls, multi-quark states, or mesonic molecules. 

Two possible non-conventional scalar mesons are the isoscalar $f_{0}$ 
and the isovector $a_{0}$. 
These particles can be studied in the $\phi$ radiative decays
$\phi \to f_{0} \gamma $ and $\phi \to a_{0} \gamma $. 
The branching ratios for these processes strongly 
depend on the meson compositions: 
$BR \sim \cal O$$(10^{-5})$ for conventional $q\bar{q}$ mesons;  
$BR \sim \cal O$$(10^{-4})$ for $qq\bar{q}\bar{q}$ mesons;  
$BR <\cal O$$(10^{-5})$ for $K\bar{K}$ molecules.
The decays $f_{0}\to\pi^{0}\pi^{0}$ and $a_{0}\to\eta\pi^{0}$ 
contribute to the radiative decays $\phi \to \pi^{0}\pi^{0}\gamma$ 
and $\phi \to \eta\pi^{0}\gamma$, respectively.
However, other contributions to these decays come from 
$\phi \to \rho\pi^{0}$ with the $\rho$ decaying to
$\pi^{0} \gamma$ and $\eta \gamma$, and from
$\phi \to \sigma(600) \gamma \to  \pi^{0}\pi^{0}\gamma$.  
The $f_{0}$ and $a_{0}$ contributions are therefore extracted by 
fitting the $\pi\pi$ and $\eta\pi$ mass spectra. 

\begin{figure}[h]
\hbox to\hsize{\hss
\includegraphics[width=\hsize]{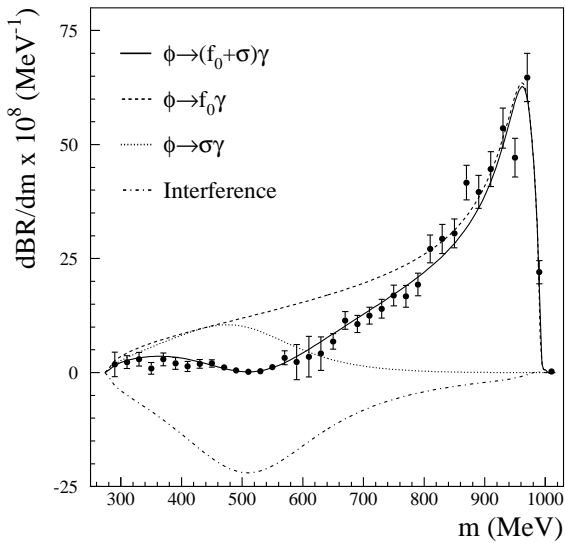}
\hss}
\caption{Mass spectra for $\phi \to \pi^{0}\pi^{0}\gamma$ decays. 
The fit is superimposed together with individual contributions.}
\label{f0}
\end{figure}

\begin{figure}[h]
\hbox to\hsize{\hss
\includegraphics[width=\hsize]{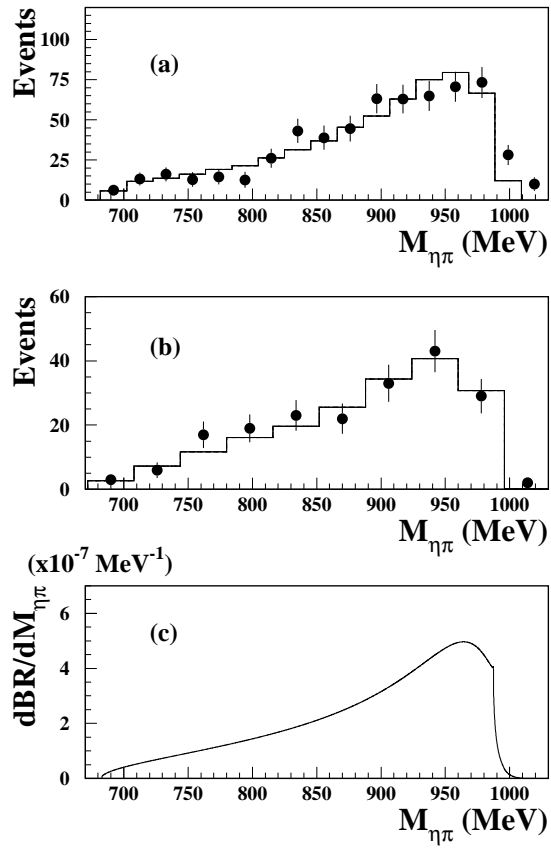}
\hss}
\caption{Mass spectra for $\phi \to \eta\pi^{0}\gamma$ decays: 
comparison of data (points) and fit (histogram) for (a) 
5 $\gamma$ final state and (b) $\pi^{+}\pi^{-} 5\gamma$ final state.
In (c) is shown the theoretical contribution of the $a_{0}$ as extracted 
from the fit.}
\label{a0}
\end{figure}

The branching ratios and mass spectra (Figs.~\ref{f0} and~\ref{a0})
for the decays $\phi \to \pi^{0}\pi^{0}\gamma$~\cite{f0kloe}  
and $\phi \to \eta \pi^{0}\gamma$~\cite{a0kloe}
have been measured by the KLOE collaboration using data 
collected in the year 2000 corresponding to an integrated luminosity of
$\sim 16~\mbox{pb}^{-1}$ ($\sim 5\times 10^{7}$ $\phi$ mesons). 
The mass spectra have been fitted using a 
theoretical framework mainly due to Achasov and 
co-workers~\cite{fitmeth1}~\cite{fitmeth2}~\cite{fitmeth3}. However,
there is no general agreement on the fit method. 
Several other authors have analyzed the KLOE data using alternative
fits~\cite{fitacha}~\cite{fitpenn}~\cite{fitoller}.   

\subsection{Event selection}

$\phi \to \pi^{0}\pi^{0}\gamma\to 5 \gamma$ events are selected first of all
by looking for five photons with total energy close to the $\phi$ mass
and total momentum close to zero. Specifically,
we require 5 clusters in the calorimeter with a time compatible
with the photon hypothesis, with total energy $E_{TOT}>800~{MeV}$ 
and total momentum $|\vec{p}_{TOT}|<200~\mbox{MeV/c}$. 
Sources of background for this decay are 
$e^{+} e^{-} \to \omega \pi^{0} \to \pi^{0}\pi^{0}\gamma$,
$\phi \to \eta \pi^{0} \gamma \to 5 \gamma$, and 
$\phi \to \eta \gamma \to 3 \pi^{0} \gamma$ decays with 
signal-to-background ratios of 0.8, 3.52, and  0.027,
respectively. Energy and momentum conservation for the five photons
are imposed in a kinematic fit in which  photons
coming from a $\pi^{0}$ are paired. The correct pairing is obtained in
$\sim 90\%$ of the time. Events with a $\pi\gamma$ invariant mass
consistent with the $\omega$ mass are identified as $\omega \pi$ events
and rejected. Data and Monte Carlo (MC) distributions are compared and good
agreement is found. The background contribution, which is obtained 
from MC simulation, is then subtracted from the 
3102 selected events. $2438\pm 61$ $\phi\to\pi^{0}\pi^{0}\gamma$
events remain. The signal acceptance $\epsilon_{\pi\pi\gamma}\sim40\%$
is obtained using MC simulation and data control samples
such as $\phi \to \eta \gamma \to 3 \pi^{0} \gamma$ events. A systematic
uncertainty of $2\%$ has been estimated. The measured branching ratio is:

\noindent
\begin{center}
\begin{equation}
BR(\phi\to\pi^{0}\pi^{0}\gamma)
       = (1.09\pm0.03_{stat}\pm0.05_{syst})\times 10^{-4}
\label{brppg}
\end{equation}
\end{center}     

The $\pi^{0}\pi^{0}$ mass spectrum is fitted by combining experimental
effects, such as mass resolution and wrong pairing, with the theoretical
function. The ratio between the function with and without experimental 
effects folded in is used to 
correct the mass spectrum. 
The purely theoretical spectrum is shown in Fig.~\ref{f0}. 
Two different final states are selected for $\phi \to \eta\pi^{0}\gamma$ 
decays: $\phi\to 5\gamma$ and $\phi\to\pi^{+}\pi^{-} 5\gamma$.
The main background processes for the $5\gamma$ final state are
$e^{+} e^{-} \to \omega \pi^{0} \to \pi^{0}\pi^{0}\gamma$,
$\phi \to \pi^{0} \pi^{0} \gamma$, and 
$\phi \to \eta \gamma$ with $\eta \to \gamma\gamma$ and $\eta\to3\pi^{0}$.
For $\phi\to\pi^{+}\pi^{-} 5\gamma$ decays there is no background
with the same final state, so only decays with similar topologies
but larger branching ratio contribute. These are 
$\phi\to\eta\gamma\to\pi^{+}\pi^{-} 3\gamma$,  
$e^{+} e^{-} \to \omega \pi^{0} \to \pi^{+}\pi^{-} 4\gamma$, and
$K_{S}K_{L}$ decays with kaons decaying into both charged and neutral pions
close to the interaction point (IP).

$\phi\to 5\gamma$ events are selected as in the previous analysis
by requiring 5 photon clusters and constraining energy-momentum conservation
with a kinematic fit. Events with a $\gamma\gamma$ invariant mass close to
the $\eta$ mass are retained. Events with a photon with energy greater than
$340$ MeV are identified as $\phi \to \eta \gamma$  and rejected.
$\gamma\gamma$ invariant masses are used to pair photons from
$\pi^{0}$'s and $\eta$'s. $\omega\pi$ and $\pi\pi\gamma$ events are rejected
using the $\pi\gamma$ and $\pi\pi$ invariant masses obtained by pairing
the photons in the two different hypotheses. The final sample consists
of 916 events. The average acceptance is $\sim 32\%$.
The number of background events as estimated from MC 
simulation is $309\pm 20$.

$\phi\to\pi^{+}\pi^{-} 5\gamma$ events are selected by requiring two tracks
of opposite charge connected to a vertex at the IP and 5
photon clusters. A kinematic fit is then performed imposing energy-momentum 
conservation and the $\eta$ and $\pi^{0}$ mass constraints. 
197 events are found with about 10 background events. The acceptance
($\sim 20\%$) obtained from MC has been corrected using 
$K_{S}\to\pi^{+}\pi^{-}$ data sample to obtain tracking efficiency, and
a $e^{+}e^{-}\gamma$ sample to obtain cluster efficiency.

For the $\phi \to \eta\pi^{0}\gamma$ branching ratio, we obtain
from the $5\gamma$ final state:
\noindent
\begin{center}
\begin{equation}
BR(\phi\to\eta\pi^{0}\gamma)
       = (8.51\pm0.51_{stat}\pm0.57_{syst})\times 10^{-5}
\label{brpega1}
\end{equation}
\end{center}     
and from the $\pi^{+}\pi^{-}5\gamma$ final state
\noindent
\begin{center}
\begin{equation}
BR(\phi\to\eta\pi^{0}\gamma)
       = (7.96\pm0.60_{stat}\pm0.40_{syst})\times 10^{-5}
\label{brpega2}
\end{equation}
\end{center}     
The mass spectra and the fit are shown in Fig.~\ref{a0}. 
 
According to the results of the fits to the $\pi^{0}\pi^{0}\gamma$ 
and $\eta\pi^{0}\gamma$ events, the $f_{0}$ and the $a_{0}$ give
the dominant contributions to these decays, and the branching ratios 
are $(4.47\pm0.21)\times 10^{-4}$ and $(7.4\pm 0.7)\times 10^{-5}$ 
for $\phi\to f_{0}\gamma$ and $\phi\to a_{0}\gamma$ decays,
respectively. According to the predictions given in the previous section
for various meson compositions, the $f_{0}$ is likely to be a  $qq\bar{q}\bar{q}$
state, while the value obtained for the $a_{0}$  doesn't distinguish between
2- and 4-quark states.

\section{$\Gamma(K_{S}\to \pi^{+}\pi^{-}(\gamma))/\Gamma(K_{S}\to \pi^{0}\pi^{0})$
  and $\delta_{0}-\delta_{2}$}

Assuming that electromagnetic interactions can be neglected,
the amplitudes for $K\to\pi\pi$ decays can be written as:
\noindent
\begin{center}
  \begin{eqnarray}
    \label{isoampl}
    A(K_{1}\to\pi^{+}\pi^{-}) & = & 
    \sqrt{2/3}A_{0} e^{i\delta_{0}} + \sqrt{1/3} A_{2} e^{i\delta_{2}}
    \\   
    \nonumber
    A(K_{1}\to\pi^{0}\pi^{0}) & = & 
    -\sqrt{1/3}A_{0} e^{i\delta_{0}} + \sqrt{2/3} A_{2} e^{i\delta_{2}}
    \\    
    \nonumber
    A(K^{+}\to\pi^{+}\pi^{0}) & = & \sqrt{3/2} A_{2} e^{i\delta_{2}}
  \end{eqnarray}
\end{center}
where $A_{0,2}$ are the amplitudes to $\pi\pi$ states with
isospin 0 and 2 respectively. 
Unitarity and CPT-invariance requires inclusion 
of the S-wave $\pi\pi$ scattering phases~\cite{derafael}
in Eq.~(\ref{isoampl}).
However, the value extracted from the previously measured
branching ratios of kaon decays $(56.7\pm 3.8)^{\circ}$ 
does not agree with the  values obtained from $\pi\pi$ 
scattering. For instance, the most recent evaluations are 
$(47.7\pm 1.5)^{\circ}$~\cite{colan} and
$(48.4\pm 2.1)^{\circ}$~\cite{yndu}. 
This disagreement is ascribed to two effects~\cite{gard}~\cite{ciri}.
The first is the presence of isospin breaking effects 
such as electromagnetic interactions. In presence of these effects
the isospin amplitudes become~\cite{ciri}:
\noindent
\begin{center}
\begin{equation}
A_{I} \Rightarrow (A_{I}+\delta A_{I})e^{i(\delta_{I}+\gamma_{I})}
\label{newampl}
\end{equation}
\end{center}     
Hence, the phase shift measured from kaon decays using Eq.~(\ref{isoampl}), 
$(\delta_{0}-\delta_{2})_{K}$,
differs from the $\pi\pi$ phase shift difference by a factor $\delta_{em}$:
\noindent
\begin{center}
\begin{equation}
  (\delta_{0}-\delta_{2})_{K}=(\delta_{0}-\delta_{2})_{\pi\pi}+\delta_{em}
\label{phaseeff}
\end{equation}
\end{center}  
An estimate of this difference is $3.2^{\circ}$~\cite{ciri}.
The second is the uncertainty on what portion of
the $\pi\pi\gamma$ decay is included in the $\pi^{+}\pi^{-}$
branching ratio. The last measurement of 
$\Gamma(K_{S}\to \pi^{+}\pi^{-}(\gamma))/\Gamma(K_{S}\to \pi^{0}\pi^{0})$ 
ratio was performed in 1976 with a 4.3\% accuracy~\cite{everhart} and 
there is no information on the procedure used to handle the radiated
photon. This uncertainty lead to a large error on the
estimate of $\delta_{0}-\delta_{2}$.

The KLOE collaboration measured this ratio using an integrated
luminosity of $17~\mbox{pb}^{-1}$ collected during the year 2000.
The KLOE measurement fully includes 
$\pi\pi\gamma$ radiative decays~\cite{ratiokloe}.

\subsection{Event selection}

\begin{figure}[h]
\hbox to\hsize{\hss
\includegraphics[width=\hsize]{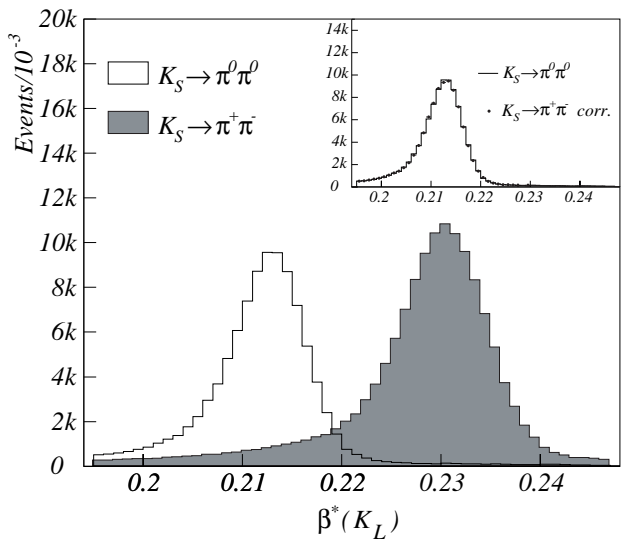}
\hss}
\caption{Distribution of the $K_{L}$ velocity $\beta^{*}$
in the $\phi$ rest frame measured from $K_{L}-crash$ clusters, for 
$K_{S}\to\pi^{0}\pi^{0}$ decays (empty histogram) and
$K_{S}\to\pi^{+}\pi^{-}$ decays (shaded histogram). In the inset,
the distribution for $K_{S}\to\pi^{+}\pi^{-}$ decays (dots), corrected for
the wrong $T_{0}$ determination, is compared with that for
$K_{S}\to\pi^{0}\pi^{0}$.}
\label{kcref}
\end{figure}

The $\phi$ decays $\sim 34\%$ of the time into $K^{0}\bar{K}^{0}$.
The decay occurs nearly at rest and the two kaons are collinear.
Since the initial state is $J^{PC}=1^{--}$, the final state
is always $K_{S}K_{L}$. This peculiarity allows us to select
a pure $K_{S}$ beam by identifying a $K_{L}$ inside the
detector ($K_{S}$-tagging). 
While the $K_{S}$ has a decay length $\lambda_{S}\sim 0.6~\mbox{cm}$ 
and decays close to the IP, the 
$K_{L}$ decay length is $\lambda_{L}\sim 350~\mbox{cm}$. Hence about
$50\%$ of the $K_{L}$ reach the calorimeter before decaying. 
The $K_{L}$ interaction in the calorimeter ($K_{L}-crash$) is
identified by requiring a cluster not associated to any track, 
with energy above 200 MeV, and velocity $\beta$ obtained by time-of-flight
compatible with the kaon velocity $\sim 0.21$. 
The $K_{L}-crash$ is used to tag the $K_{S}$.
The correct measurement of the $K_{L}$ velocity ($\beta^{*}$) requires
the subtraction of the initial time of the event $T_{0}$ from
the cluster times. Due to the small bunch crossing period, 
$\sim 3~\mbox{ns}$, the $T_{0}$ in KLOE has to be determined event by event. 
A first estimate is obtained by assuming that the
first cluster to arrive, with $E_{cl}>50~\mbox{MeV}$ and 
distance to the beam line $\rho_{cl}>60~\mbox{cm}$ is due to
a photon. This assumption, which is correct for $K_{S}\to\pi^{0}\pi^{0}$
but not for $K_{S}\to\pi^{+}\pi^{-}$, leads to a different $\beta^{*}$
distribution for the two decay modes as shown in Fig.~\ref{kcref}.
Therefore, different tagging efficiencies are obtained if
$K_{L}-crash$ cluster is selected  
by requiring $0.195<\beta^{*}<0.2475$.
The correction is obtained by measuring 
the $T_{0}$-corrected distribution for $K_{S}\to\pi^{+}\pi^{-}$ 
from a control sample and comparing it with the uncorrected distribution. 

$K_{S}\to\pi^{0}\pi^{0}$ events are selected by requiring at least 3
photons defined as clusters not associated to any track 
with $|t_{cl}-r_{cl}/c|< 5 \sigma_{t}$. The background is rejected
by cutting on energy and polar angle: $E_{cl}>20~\mbox{MeV}$
and $\cos{\theta_{cl}}<0.9$. The distribution of the number of
clusters in $\pi^{0}\pi^{0}$ is shown in Tab.~\ref{tab:2pi0}. 
Almost 90\% of the events are retained.
The effects of cluster splitting and accidental
clusters are reduced by an order of magnitude by these cuts.
The photon detection efficiency is measured from
$\phi\to\pi^{+}\pi^{-}\pi^{0}$ samples and used to correct
the MC simulation.   

\begin{center}
\begin{table}[h]
\caption{Distribution of the number of clusters in $\pi^{0}\pi^{0}$ events.}
\label{tab:2pi0}
{\footnotesize
\begin{center}
\begin{tabular}{|c|r|}
\hline
{} &{} \\[-1.5ex]
N. clusters & \%  \\[1ex]
\hline
 {} &{}  \\[-1.5ex]
 $<$ 2  & 1.3  \\[1ex]
 2  & 8.6  \\[1ex]
 3  & 33.2 \\[1ex]
 4  & 56.6 \\[1ex]
 $>$ 4 & 0.3  \\[1ex]
\hline
\end{tabular}
\end{center}
}
\vspace*{-13pt}
\end{table}
\end{center}

\begin{figure}[h]
\hbox to\hsize{\hss
\includegraphics[width=\hsize]{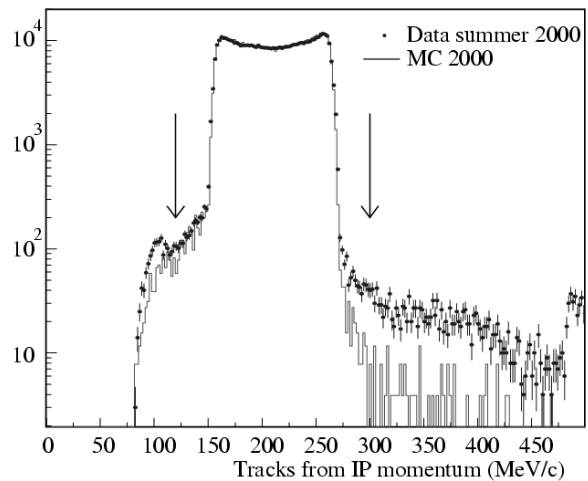}
\hss}
\caption{Momentum distribution for tracks in tagged events.
Data (dots) and MC (histogram). The arrows show the cut applied.}
\label{tracks}
\end{figure}

\begin{figure}[h]
\hbox to\hsize{\hss
\includegraphics[width=\hsize]{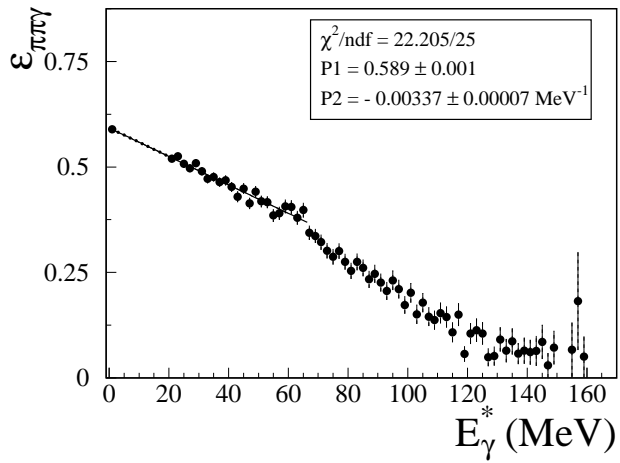}
\hss}
\caption{Acceptance for $K_{S}\to\pi^{+}\pi^{-}\gamma$ 
events as a function of the energy of the radiated photon,
obtained from MC $\pi^{+}\pi^{-}$ simulation ($E_{\gamma}=0$)
and from MC $\pi^{+}\pi^{-}\gamma$ simulation 
($E_{\gamma}>20~\mbox{MeV}$).}
\label{kpprad}
\end{figure}

$K_{S}\to\pi^{+}\pi^{-}(\gamma)$ events are selected by requiring
the presence of two tracks of opposite charge coming
from the IP that reach the calorimeter.
The momentum distribution for charged
tracks in tagged events is shown in Fig.~\ref{tracks}. 
A peak due to charged kaons is visible at 100 MeV in the data (dots)
as well as a long tail at higher momentum due to residual background. 
These events are rejected by requiring $120<p<300~\mbox{MeV/c}$.
The efficiency for reconstructing a track is measured from
a control sample of $K_{S}\to\pi^{+}\pi^{-}$ decays and then combined
with the MC simulation. The overall acceptance is obtained
from MC taking the radiated photon into account. The acceptance
is shown in Fig.~\ref{kpprad} as a function of the photon
energy. The value for $E_{\gamma}=0$ is obtained from
$\pi^{+}\pi^{-}$ simulation, and for $E_{\gamma}>20~\mbox{MeV}$
from $\pi^{+}\pi^{-}\gamma$ simulation. The acceptance
in the region $0<E_{\gamma}<20~\mbox{MeV}$ is obtained by
linear interpolation.
The decrease in acceptance is due to the requirement
that both pions reach the calorimeter. 
For high values of the photon energy the
probability for both pions to reach the calorimeter is low.
The overall acceptance is obtained
by folding this efficiency with the photon spectrum from~\cite{ciri}.
The inclusion of the radiated photon leads to a correction
of $\sim0.3\%$.

In the data collected during the year 2000 we identify
1,060,821 $K_{S}\to\pi^{+}\pi^{-}$ and
766,308  $K_{S}\to\pi^{0}\pi^{0}$ events.
The ratio of the branching ratios is obtained by correcting
the ratio of these numbers with the acceptance and
the tagging efficiencies.
The result is:
\noindent
\begin{center}
\begin{equation}
\frac{BR(K_{S}\to\pi^{+}\pi^{-}(\gamma))}{BR(K_{S}\to\pi^{0}\pi^{0})}
       = (2.236\pm0.003_{stat}\pm0.015_{syst})
\label{ratioresult}
\end{equation}
\end{center}  
This value is higher than the world average, $2.197\pm0.026$, as we
expect due to the inclusion of all the radiative decays.
The error, $0.7\%$, is dominated by
systematics. The statistical error is already
at the $0.1\%$ level. The main contribution to the systematic
error comes from the correction for the different tagging
efficiencies. The cut on $\beta^{*}$ was loosened
in the year 2001 in order to reduce the contribution from
this correction.

Using this new value, the phase shift
is found to be:
\noindent
\begin{center}
\begin{equation}
(\delta_{0}-\delta_{2})_{K}=(47.8\pm2.8)^{\circ}
\label{newdiff}
\end{equation}
\end{center}  
which is now in agreement with the values extracted 
from $\pi\pi$ scattering. The contribution of the electromagnetic
interactions can be estimated from the difference between 
$(\delta_{0}-\delta_{2})_{K}$ and $(\delta_{0}-\delta_{2})_{\pi\pi}$.
For instance, using the value in~\cite{colan}:
\noindent
\begin{center}
\begin{equation}
(\delta_{0}-\delta_{2})_{\pi\pi}-(\delta_{0}-\delta_{2})_{K} =(-0.1\pm3.2)^{\circ}
\label{emcont}
\end{equation}
\end{center}

\section{Conclusion and perspectives}

With the data collected during the year 2000, 
the KLOE collaboration has improved the previous existing
measurements of branching ratios and mass spectra
for $\phi$ radiative decays and the ratio 
$BR(K_{S}\to\pi^{+}\pi^{-}(\gamma))/BR(K_{S}\to\pi^{0}\pi^{0})$.
After two more years of data taking, the total integrated luminosity
has increased by a factor of $\sim 20$, allowing new and more
precise measurements.
Preliminary results on radiative decays are in agreement
with the previous ones. New analyses are also underway in order
to study the  Dalitz plot for 
$\phi\to\pi^{0}\pi^{0}\gamma$ decays and the branching ratio
and mass spectrum for $\phi\to\pi^{+}\pi^{-}\gamma$ decays
in order to look for the $f_{0}\gamma$ intermediate state. 

The ratio $BR(K_{S}\to\pi^{+}\pi^{-}(\gamma))/BR(K_{S}\to\pi^{0}\pi^{0})$
has a statistical error which is already at per-mil level while
the systematic error is $\sim 0.7\%$.
However, the higher statistics, together with the looser cut on the
$\beta^{*}$ distribution should allow us to reduce 
the systematic error. 
With the expected $\sim 0.1\%$ fractional error, we can measure
$(\delta_{0}-\delta_{2})_{K}$ with an error of $\sim 0.5^{\circ}$.
Finally we are also planning to measure the energy distribution
of the radiated photon.

\end{document}